# Efficient and Robust Secure Aggregation for Sensor Networks

Parisa Haghani, Panos Papadimitratos, Marcin Poturalski, Karl Aberer, Jean-Pierre Hubaux
Email: {firstname.lastname@epfl.ch}

*Abstract*—Wireless Sensor Networks (WSNs) rely on in-network aggregation for efficiency, however, this comes at a price: A single adversary can severely influence the outcome by contributing an arbitrary partial aggregate value. Secure in-network aggregation can detect such manipulation [2]. But as long as such faults persist, no aggregation result can be obtained. In contrast, the collection of individual sensor node values is robust and solves the problem of availability, yet in an inefficient way. Our work seeks to bridge this gap in secure data collection: We propose a system that enhances availability with an efficiency close to that of in-network aggregation. To achieve this, our scheme relies on costly operations to localize and exclude nodes that manipulate the aggregation, but *only* when a failure is detected. The detection of aggregation disruptions and the removal of faulty nodes provides robustness. At the same time, after removing faulty nodes, the WSN can enjoy low cost (secure) aggregation. Thus, the high exclusion cost is amortized, and efficiency increases.

## I. Introduction

Wireless sensor networks (WSNs) have evolved to a valuable tool for numerous applications. The rationale of current and future WSNs is to deploy a multi-hop wireless network of low-complexity and low-cost sensor nodes, with each of them able to operate for a long period of time with a single energy source. Depending on the application, sensors either report each and every measurement to a *Base Station (BS)* or *sink*, or they perform *in-network aggregation*: *En route* to the sink, nodes combine their own measurement with the ones of other nodes in proximity, e.g., their children on an *aggregation tree* rooted at the sink and spanning all sensors. A large fraction of WSNs requires only a periodic collection of an aggregate value (e.g., count, sum, average), and can do so with low network overhead. With in-network aggregation, rather than relaying individual measurements across multiple hops, each node transmits a single packet, "summarizing" the data from an entire area of the WSN, e.g., the node's aggregation subtree.

However, in-network aggregation is a two-edged sword, making compromise of the collected data much easier. Clearly, a faulty or adversarial node can always inject erroneously its own sensory data, which, if treated properly, may have little effect on the network-wide data [11]. But by forwarding a false aggregate value, an adversarial node can severely affect the overall data aggregate. Consider, for example, aggregation over a tree and an adversarial node $x$ being a neighbor of the sink. $x$ can control the data contributions of a significant fraction of the network.

It is thus critical to safeguard the WSN operation against such attacks. To secure data collection, the simplest approach would be for each node to authenticate its measurements to the sink in an end-to-end manner. But in that way, all measurements would need to be transmitted individually: This *elementary* approach abolishes the reduction of network overhead, the advantage of in-network aggregation.

The benefit and vulnerability, as well as the need to secure in-network aggregation, have been identified by a number of schemes in the literature. One approach relies on the placement of adversarial nodes and correct nodes monitoring their parents [13]. This implies that a few compromised neighboring nodes could still abuse the system. Another approach proposes a probabilistic check of parts of the aggregation and use of a presumed data structure to detect outliers [14]. Nonetheless, attacks could be undetected with some probability, or be non-detectable altogether, if a data model is not available or applicable. This is why in this work we are after a system that is generally applicable and robust to any configuration of the adversary.

In this direction, the *Secure Hierarchical In-network Aggregation (SHIA)* scheme [2] achieves detection of any manipulation of the in-network aggregation, without assuming a particular data structure (e.g. correlation). In short, SHIA guarantees *optimal security* [2], which means that the adversary is allowed to modify values (sensor data) of the nodes it controls, with the sink accepting an aggregation result if and only if the values of all non-faulty nodes were properly aggregated. The efficiency of SHIA stems from the fact that nodes themselves verify and acknowledge the correctness of the aggregation. The acknowledgements are then delivered to the BS in such a way, that it is possible to detect if all nodes acknowledged: So manipulations of the aggregation are always detected, yet SHIA cannot but *reject* the result of a manipulated aggregation. In other words, it suffices to have a single adversary present, disturbing each aggregation, thus *denying* access to any measurement from the sensor field.

One solution to this problem, stated by the Chan, Perrig and Song [2], is to use the elementary scheme as a fall-back solution in case SHIA fails. This would ensure availability but at a high cost. In fact, such a solution implies that, in the presence of faulty nodes, the efficiency of in-network aggregation is *not* utilized. This is exactly the problem we solve in this paper: *How to achieve both efficiency and availability for data collection in sensor networks*.

We propose a system that relies on SHIA for efficient in-network aggregation and detection of any manipulation. Upon detection, we invoke more expensive protocols that localize and exclude nodes deemed faulty. Our protocols deliver the SHIA acknowledgments to the BS using "onion" authentication, enabling identification of non-acknowledging nodes and successively localizing those misbehaving with only low uncertainty. This way, and without relying on any assumptions on the data model or the faulty nodes configuration, our system can revert to efficient in-network aggregation after the exclusion of faulty nodes, and thus perform numerous cost-effective data extractions, rather than a series of expensive elementary data collections. To the best of our knowledge, this is the first work to provide this generalization on secure in-network aggregation, that is, state the requirement for robust, highly available and at the same time efficient data collection. Our solution to this problem, specified in Sec. II, is presented in Sec. III and analyzed in Sec. IV, before a discussion on related and future work and conclusions.

## II. ASSUMPTIONS, MODEL, AND SPECIFICATION

### A. System Model

We consider a WSN comprising $n$ sensor nodes, each with a unique identifier $s$, and a single base station (BS). We assume that the network is well connected, i.e. every node has a considerable number of neighbors, at most $d_{max}$. Each node $s$ shares a symmetric key $K_s$ with the BS, and one symmetric key $K_{s,t}$ with each neighbor $t$. These keys are either preloaded or established at deployment time using one of the methods in the literature, e.g., [3]. Communication between neighbors $s$ and $t$ is always authenticated using $K_{s,t}$. A data link broadcast/multicast from $s$ can be authenticated using the $K_{s,v_1}, \ldots, K_{s,v_k}$ keys shared with the $v_i$ neighbors. Unless noted otherwise, such a *local authenticated broadcast* is used. In addition, the BS can perform a symmetric-key based *network-wide authenticated broadcast* using a protocol such as $\mu$TESLA [8].

The goal of the WSN is to calculate an aggregate $agr$ (e.g. sum, mean) of node measurements, in a process we call an *aggregation session* or, for simplicity, aggregation. Each aggregation $A$ is identified by a unique session identifier $N_A$, selected by the BS. An aggregation is terminated by the BS, which declares it either *successful*, with an *aggregation value* $val_A$ returned, or *failed* and no value returned.

The node measurements lie in a range $M$; we refer to each $val_A(s) \in M$ as the node's *value* and consider $val_A(s)$, in general, independent of any $val_A(t)$ for any $t \neq s$ during the same aggregation $A$. Data extraction is performed across the *aggregation tree* $(T_A)$ rooted at the BS; for simplicity, the BS has a single child. It also knows the whole $T_A$, whereas nodes know their parent and children in $T_A$. In Sec. III-D we propose a tree construction protocol with the robustness sought for our scheme. We can schedule the node actions during aggregation according to the method in [4], with children of a node responding before it does. This method is used both by SHIA and our protocols. For compliance with other works, we assume that an aggregation tree $T_A$ is in place for the first aggregation. If $A$ is deemed failed, a new aggregation tree $T'_A$ is constructed, to be used in the subsequent aggregation session, which does not include the nodes deemed faulty. The *cost* of an aggregation is defined as the maximum edge congestion, that is, the number of bytes transmitted over a link due to the protocol activity.

### B. Adversary Model

We assume that the adversary can compromise sensor nodes, obtaining their cryptographic keys and controlling their functionality; it can thus induce arbitrary deviations from the protocol, even in a coordinated manner among compromised nodes. We refer to nodes that deviate from the protocol (including benign failures) as *faulty* nodes, and nodes that do not as *correct*. However, the contribution of an arbitrary own value $val_A(s)$ different than its actual measurement is not considered as a deviation: Without any assumptions on data correlations, $s$ is the only node responsible and able to perform this measurement. We assume that faulty nodes are aware of the tree $T_A$, and that the BS is always correct (i.e., cannot be compromised by the adversary).

We are not concerned with jamming and denial of service (DoS) in various protocol layers [5], [12], Sybil/Node replication attacks [6], [7] or "wormhole" formation [1], [10]; these attacks are beyond the scope of this work, and countermeasures against them can coexist with our protocols. However, notification of the BS upon detection of a jamming attack by correct nodes is needed; otherwise, a jammed node could be unnecessarily excluded by the BS.

### C. Specification

We are interested in protocols that can perform a sequence of aggregations $\{A_1, A_2, \ldots, A_j\}$ in a WSN with $n_a$ faulty nodes, and satisfy the following properties:

**Security:**
- At most $n_a$ aggregations fail.
- For every successful aggregation $A \in \{A_1, A_2, \ldots A_j\}$, with $V_A$ being a multi-set of values contributed by correct nodes in the aggregation tree $T_A$, and $V'_A$ a multi-set of arbitrary values in range M equal in size to the number of faulty nodes in $T_A$, it holds that
$$val_A = agr(V_A + V'_A)$$

**Efficiency:**
- At most $n_a$ aggregations have communication complexity of $O(n)$.
- For any other aggregation $A$, the cost is $O(h_A \Delta_A)$, with $h_A$ the height and $\Delta_A$ is the maximum node degree for $T_A$.

This specification allows the adversary to disrupt at most $n_a$ aggregation sessions, which is at most one per faulty node. A disrupted aggregation is allowed to be relatively expensive and not produce an aggregation value – intuitively, it is sacrificed to cope with the faulty nodes that caused the disturbance. A successful aggregation, however, should output a result complying with *optimal security* [2] and should also be relatively inexpensive.

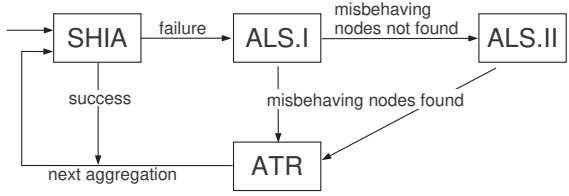

Fig. 1. Scheme overview

## III. PROPOSED SCHEME

In this section, we present our solution, illustrated in Fig. 1. The system operates in three stages. First, data aggregation and manipulation detection are performed by SHIA [2], due to its efficiency and effectiveness. If successful, the system proceeds to the next aggregation. Otherwise, at a second stage, the *Adversary Localizer Scheme (ALS)* is launched: the ALS.I phase localizes, i.e., *marks*, nodes that disrupted the aggregation value, and ALS.II marks nodes that disrupted the acknowledgement collection during stage one (SHIA). At the third stage, the *Aggregation Tree Reconstruction (ATR)* protocol is invoked, which constructs a new aggregation tree excluding the marked nodes. Gradually, after a series of failed aggregations, all the faulty nodes will be excluded, allowing undisrupted in-network aggregation and thus efficient operation.

Below, we use the following cryptographic primitives: $H$, a collision-resistant hash function, and $MAC$, a Message Authentication Code. $Auth_K(m)$ denotes message $m$ authenticated using the symmetric key $K$, e.g. $<m, MAC(m)_K>$.

### A. Stage One: SHIA

We present the *naive* version of SHIA in detail, as our ALS protocol relies on its functionality rather than using it simply as a "black-box". ALS can be also adapted to the *improved* SHIA version. The SHIA algorithm focuses on the *sum* aggregate and consists of three phases: *query dissemination*, *aggregate-commit* and *result checking*. The base station (BS) initiates the aggregation, generating a nonce $N$ that identifies the aggregation session and broadcasting it to the network, as part of a *query* (along with other possibly useful data) in an authenticated manner.

Then, in the *aggregate-commit* phase, every node calculates a *label*, based on the labels of its children and its own value, and sends it to its parent node. The label is a <count, value, commitment> tuple, with count the number of nodes in the subtree rooted at the node, value the sum of all the nodes values in the subtree, and commitment the cryptographic commitment tree over the data values and the aggregation process in the subtree.[1] For a leaf node $s$, the label has the format: $<1, val(s), s>$. For an internal node $t$, suppose its children have the following labels, $l_1, l_2, ..., l_q$, where $l_i = <c_i, v_i, h_i>$; among these is also $t$'s value, formatted as a leaf node's label: $<1, val(t), t>$. Then, the label of $t$ is $<c, v, h>$, with $c = \sum c_i$, $v = \sum v_i$, and $h = H[N\|c\|v\|l_1\|l_2\|...\|l_q]$. Nodes store the labels of their children, as they will be used later on. This phase ends with the BS receiving the label $<c, v, h>$ of its single child $b$. The aggregate $v$ will be declared as the aggregation value if the result-checking phase, described next, is successful.

In the *result-checking* phase, the BS disseminates, using an authenticated broadcast, $N$ and the $<c, v, h>$ label. Every node uses this label to verify if its value was aggregated correctly. To do this, each node $s$ is provided with the labels of its *off-path* nodes, that is, the set of all the siblings of the nodes on the path from $s$ to the root of the tree. These are forwarded across the aggregation tree: a parent $t$ provides every child $s$ with the labels of $s$'s siblings (which it stored during the previous phase), along with every off-path label received from its parent.

With all off-path labels at hand, $s$ recomputes the labels of all its ancestors in the aggregation tree all the way to the root, and compares the result to $<c, v, h>$ provided by the BS. As proved in [2], they match only if all the nodes on the path from $s$ to the root aggregated $s$'s value correctly (although this does not guarantee that the values of other nodes were not mistreated). If so, $s$ *acknowledges* the result. We observe here the following simple fact:

*Fact 1:* If a node and its child both follow the SHIA protocol, either they both acknowledge or neither do.

A node $s$ acknowledges by releasing an *authentication code (ack)*: $MAC_{K_s}(N\|OK)$, where OK is a unique message identifier and $K_s$ is the key shared between node $s$ and the BS. Leaf nodes send their *ack* while intermediate nodes wait for *ack*s from all their children, compute the XOR of those *ack*s with their own *ack*, and forward the resultant *aggregated ack* to their parent.

Once the BS has received the aggregated ack message $A_b$ from $b$, it can verify whether all nodes acknowledged the aggregation value: It calculates the *ack* of every sensor (using the key shared with the node), XOR'es them and compares the result to $A_b$. In case of equality, all nodes acknowledged, and the BS declares the aggregation successful. Otherwise, our ALS protocol is triggered.

### B. Stage Two: Adversary Localizer Scheme, Part I

ALS.I marks nodes that misbehaved in the *aggregate-commit* phase of SHIA or the dissemination of *off-path* values. ALS.I consists of two phases:

*1) Hierarchical Collection of Confirmations:* The BS initiates this phase by sending an authenticated broadcast containing $N$ and informing all nodes that ALS.I is taking place. If a node had not acknowledged the result (as determined by SHIA), it does not respond. Otherwise, a leaf node $s$ sends a confirmation $M_s = Auth_{K_s}(N)$ to its parent. An internal node $t$ waits for its children, $u_1, u_2, ..., u_k$ (the order is based on their identifiers) to send their confirmations. Then, $t$ sends up the following confirmation: $M_t = Auth_{K_t}(N, M_{u_1}, M_{u_2}, ..., M_{u_k})$. If $t$ does not receive any messages from its $r^{th}$ child, $M_{u_r}$ is replaced with a predefined message $M_{nr}$, indicating "no message received from this child". See Fig. 2 for illustration.

---
[1] For brevity, we omit the complement field, for details see [2].

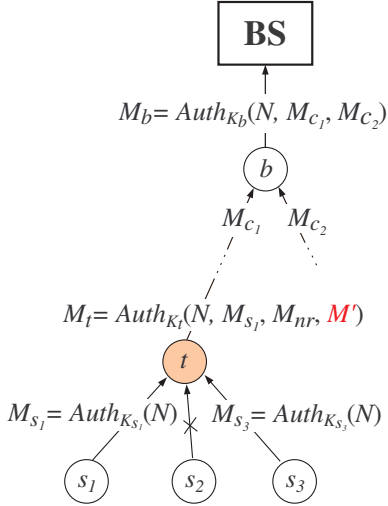

Fig. 2. ALS.I: Hierarchical Collection of Confirmations. Nodes $s_1$ and $s_3$ acknowledge, node $s_2$ does not. Note $t$ is faulty: it encapsulates $M_{s_1}$, and $M_{nr}$ correctly, but modifies the message of $s_3$. The collection continues, until node $b$ provides the final confirmation to the BS.

*2) Recursive Processing of Confirmations:* The above procedure results in message $M_b$ reaching the BS. If no message is received, then $b$, the single child of the BS, is marked. Otherwise, the message is processed in a recursive manner. As the aggregation tree is known to the BS, it knows that $M_b$ should be authenticated using $K_b$. If it is, and the message begins with $N$, and the proper number of *child* messages (equal to the number of $b$'s children) can be extracted from it, the message is regarded as *legitimate*, and the recursive procedure is applied to each child message. Otherwise the message is regarded as *illegitimate*. Note that the special $M_{nr}$ message is also regarded as illegitimate. In that case, the BS marks the node to which this message corresponds to and its parent (in the farther recursive executions, when the BS is not the parent); the recursive execution stops. The recursive procedure also stops when it reaches a leaf node. See Fig. 3 for an illustration.

### C. Stage Two: Adversary Localizer Scheme, Part II

It is possible that ALS.I does not localize any faulty nodes, even though SHIA declared a failed aggregation. This can happen when all correct nodes acknowledge, which implies a correctly done aggregation but a faulty node disrupting the aggregation of *ack*s. On the positive side, in such a situation the BS can be sure of a correct aggregation result. However, the not removed faulty node could disrupt a subsequent aggregation, something unacceptable according to our problem statement.

The ALS.II scheme addresses this problem. To implement ALS.II, we need to slightly modify SHIA, making every node store the *ack* messages it received from its children. ALS.II delivers them to the BS, using the same mechanism as the Hierarchical Collection of Confirmations in ALS.I. Then, the BS identifies inconsistencies, through an algorithm based on Recursive Processing of Confirmations.

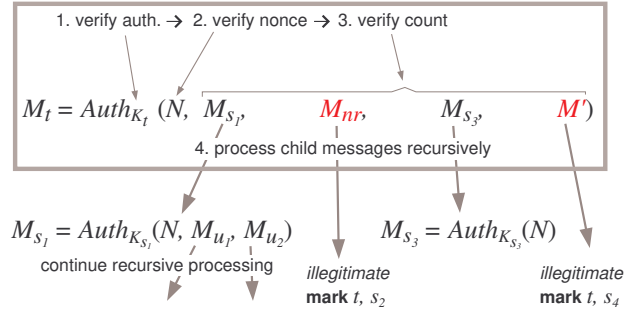

Fig. 3. ALS.I: Recursive Processing of Confirmations: recursive procedure steps, explained in the box, and unfolding of the recursion.

As we show in Lemma 1 of Sec. IV, if ALS.II is triggered, all (correct) nodes acknowledge the aggregation result. Under this assumption, the topology-aware BS can calculate for every node $s$ what the *ack* message sent by this node should be: We denote this value $\overline{A}_s$.

*1) Hierarchical Collection of acks:* The BS initiates this phase by broadcasting an authenticated message containing $N$, informing the nodes that ALS.II is taking place. A leaf node $s$ does not send anything. An internal node $t$, with non-leaf children $u_1, u_2, ..., u_q$, sends up $M_t = Auth_{K_u}(N, M_{u_1}, M_{u_2}, ..., M_{u_q}, A_{u_1}, ..., A_{u_q})$, where $M_{u_1}, M_{u_2}, ..., M_{u_q}$ are the messages $t$ received from its children in this phase, and $A_{u_1}, A_{u_2}, ..., A_{u_q}$ are *ack* messages that it received from them in the SHIA result-check phase.

*2) Recursive Processing and Ack Analysis:* Upon receiving $M_b$ from its child $b$, the BS recursively processes it to find the source(s) of discrepancy. As in ALS.I, if the message of node $t$ illegitimate, meaning that it is not authenticated with $K_b$, it does not begin with $N$, or the proper number (of nonleaf children). of child messages and *ack* messages cannot be extracted from it, both $t$ and its parent are marked. There are only two significant differences from ALS.I. First, if for some node $u$ the *ack* message $A_u$ equals $\overline{A}_u$, then the corresponding child message $M_u$ is not further processed. Second, the BS is looking for *ack inconsistencies*, which have two variations: (i) for node $t$ which has a leaf node $s$ as a child, $A_s$ is different from $MAC_{K_s}(N\|OK)$, the *ack* of node $s$ (Fig. 4a); (ii) for node $t$ which has a child $s$, which has the children $u_1, ..., u_q$, the value $A_s$ is not equal to $MAC_{K_s}(N\|OK) \otimes (\bigotimes A_{u_i})$ (Fig. 4b). If an ack inconsistency is detected, both $t$ and $s$ are marked, but the recursive procedure is continued (if possible).

### D. Stage Three: Aggregation Tree Reconstruction

The Aggregation Tree (Re)Construction (ATR) protocol, in addition to the tree construction, allows the BS to exclude nodes in $BL$, a *black list*, from the new tree $T'_A$ and provides the BS with the knowledge of $T'_A$. The primary *requirement* for ATR is that its output, $T'_A$, is identical to the parts of the tree nodes know. This way, a faulty node is not be able to mislead the ALS protocols marking a correct node.

Due to space limitations, we do not describe ATR in full detail here. We present ATR in a basic form that achieves

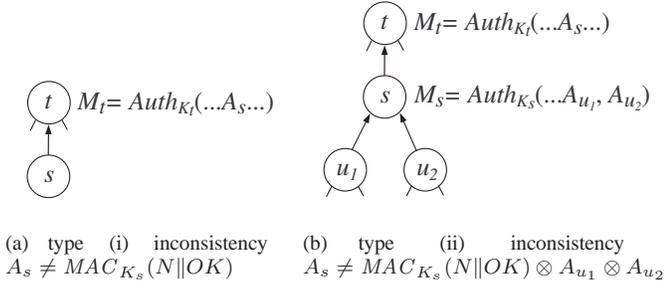

(a) type (i) inconsistency  
$A_s \neq MAC_{K_s}(N\|OK)$

(b) type (ii) inconsistency  
$A_s \neq MAC_{K_s}(N\|OK) \otimes A_{u_1} \otimes A_{u_2}$

Fig. 4. ALS.II: ack inconsistencies

our system objectives. More advanced and efficient variants are left for future work. The BS initiates ATR, via a neighboring node $b$, by sending a *tree establishment (TE)* message $<N, BL, n>$, protected by a network-wide broadcast authenticator $Auth_{Bcast}$ [8].

As the *TE* message is flooded, it is authenticated in a hop by hop manner by data-link broadcast authentication. Each $v$ maintains $s$ from which it first receives a fresh *TE* as its parent, and confirms to $s$ that it is its child. After $s$ hears from its $k$ children, namely $v_1, \ldots, v_k$, it sends its *response to the BS*: $Auth_{K_s}(N, s, (v_1, \ldots, v_k))$, or $Auth_{K_s}(N, s)$, if it is a leaf. The responses are propagated upwards to the BS. After sending its own response, the node acts as a relay for the responses of its children and up to $n$ responses or until the response collection concludes; these constraints are added to keep the cost bounded, but might result in loss of legitimate responses. A faulty node cannot create any inconsistency between the tree at BS and the nodes if responses are lost or dropped. Even if the faulty node eliminated a subtree, it would at most prevent aggregation from a part of the network, but no correct node would be blacklisted and thus permanently excluded.

*1) Highly Resilient ATR:* To ensure that the new tree ATR covers all nodes, we sketch here a different protocol, whose initial phase must run before any aggregation takes place. After each node $s$ ran a secure neighbor discovery, it floods its *neighbor list ($NL_s$)* across the network; a fresh $NL_s$ is relayed by each node only once. Upon receipt of the neighbor lists from all nodes, the BS constructs the network connectivity graph, rejecting links not announced by both neighbors. The BS then calculates locally $T_A$. At the end of this initial phase, as well as after any subsequent reconstruction, the BS simply distributes the newly calculated aggregation tree across the network. It suffices that each node relays the message containing $T_A$ once at most.

We emphasize that the costly *NL* collection is performed in general *only once*, at the initial phase of ATR. To ensure resilience to DoS attacks, nodes need to authenticate each *NL* they relay. Otherwise, faulty nodes could flood the network with bogus neighbor lists. To achieve this, public key cryptography is needed; recent implementations, e.g., see [9] and references within, attest to its feasibility for WSNs. Each responding $s$ signs its $NL_s$. The scheme cannot be exploited by clogging/energy consumption DoS attacks: Correct nodes immediately ignore messages coming from a neighbor that forwarded one invalid-signed *NL*, as the forwarder should have checked its validity already.

## IV. ANALYSIS

In this section, we show that the proposed scheme satisfies the specification. We will use the following notions: a node *misbehaves* in an aggregation session if it does not follow the protocol. Otherwise, it *behaves correctly* in an aggregation session. A correct node by definition behaves correctly all the time, where as a faulty node does not always misbehave.

*Lemma 1:* If some correct node does not acknowledge in the SHIA result-checking phase, then ALS.I marks at least one misbehaving node.

*Proof:* First, we show that some node will be marked. As some correct node $s$ does not send a confirmation message, if the recursive procedure reaches the point of evaluating the legitimateness of $M_s$, both $s$ and its parent will be marked, because it is not possible for the adversary to forge $M_s$. On the other hand, if the BS stops the recursive procedure at some ancestor of $s$, then this ancestor is marked.

Next, recall that nodes are always marked in pairs: child $s$ and parent $t$.[2] This happens if $M_s$, the confirmation message of $s$ as extracted from $M_t$, is illegitimate. We show that at least one of these nodes is misbehaving. Note first, that as $M_s$ is extracted from $M_t$, node $t$ has acknowledged. Consider two cases:

(i) Node $s$ is behaving correctly. Either $s$ sent up a legitimate confirmation and $t$ has modified it in $M_t$ (misbehavior). Or, $s$ did not send a confirmation message, which means it is not acknowledging. As $t$ is acknowledging, we can conclude from Fact 1 that $t$ is misbehaving.

(ii) Node $t$ is behaving correctly. Then either $s$ sent to $t$ an illegitimate confirmation message (misbehavior), or $t$ did not receive a confirmation from $s$, which means $s$ is not acknowledging. We use Fact 1 again to conclude that $s$ is misbehaving.

Note: Due to adversarial behavior, the aggregation, at stage 1 (SHIA), may take place over a tree different from $T_A$, the one the BS is aware of and utilizes for ALS. However, as all neighbor communications are authenticated, and a correct node is aware of its parent and children in $T_A$, the only difference between these trees can be a faulty node changing its parent to some other faulty node. This allows us to apply Fact 1. ∎

*Lemma 2:* If ALS.I is executed and it does not mark any nodes, then ALS.II marks at least one misbehaving node.

*Proof:* Based on Lemma 1, we can assume that all correct nodes acknowledge the SHIA result.

First, if a node $t$ and its child $s$ are marked because of an illegitimate message of $s$, then one of them is misbehaving. Indeed, if $s$ is behaving correctly it must have sent a legitimate message to $t$, and $t$ had to misbehave by modifying it. If $t$ is behaving correctly, then the message it received from $s$ must be illegitimate, which is only possible if $s$ is misbehaving.

---

[2]The only exception is when $t$ is the BS and only case (ii) of the proof needs to be considered.

Second, if an ack inconsistency is found, then one of the marked nodes $s$ or $t$ has to misbehave. Indeed, if the inconsistency is of type (i), either the leaf $s$ has "acknowledged" with an incorrect *ack* or not at all, which is misbehaving because all nodes should acknowledge, or $t$ is reporting $A_s$ different from what it actually received from $s$. If the inconsistency is of type (ii), then either $s$ did not send to $t$ the aggregated *ack* it should have calculated (based on what it received form its children and its own *ack*), or $t$ is reporting $A_s$ different from what it received from $s$.

If all the messages are legitimate, and no ack inconsistencies are found, a simple induction on the structure of the aggregation tree shows that, for $b$, the single child of the BS, $A_b$, the aggregated *ack* that the BS received in SHIA, is equal to $\overline{A_b}$, which is a contradiction with the fact that SHIA failed and ALS.I is executed. ∎

*Theorem 1:* Our scheme satisfies the specification.

*Proof:* First we prove that at most $n_a$ aggregations fail. If an aggregation fails, ALS is executed, and by Lemmas 1 and 2, at least one misbehaving and thus faulty node is marked and removed from the aggregation tree in the ATR phase. Thus, after at most $n_a$ failures, only correct nodes remain in the aggregation tree, and all subsequent aggregation sessions will be successful.

**Security**. If an aggregation is successful, then SHIA has not detected any misbehavior. We can thus refer to theorem 13 of [2] to obtain the desired condition on $val_A$.

**Efficiency**. In successful aggregations, only SHIA is executed, and we can refer to the *naive* counterpart of theorem 15 from [2] to get the desired cost $O(h_A \Delta_A)$. In a failed aggregation session, ALS is executed and the tree is reconstructed. A simple inductive argument shows that the cost of ALS is $O(n)$. Indeed, in both ALS.I and ALS.II, each node $u$ sends up to its parent a message that has size $O(|T_u|)$, where $T_u$ is subtree rooted at $u$. The cost of ATR comprises the request, of size $O(d_{max})$ and the forwarding of at most $n$ responses each of size $O(d_{max})$; thus $O(n)$. The cost for the highly resilient ATR is the forwarding of at most $n$ *NL*s each of size $O(d_{max})$. Thus, *NL* collection cost $O(n)$. The $T_A$ message size is $O(n)$, forwarded once; thus, $O(n)$ overall. ∎

## V. Discussion and Conclusion

Our scheme builds on the *Secure Hierarchical In-Network Aggregation* [2], in order to achieve not only secure but also efficient WSN data collection over a series of aggregations. We have described a basic version of our scheme, sufficient to satisfy the stated specification. However, there are a number of enhancements and extensions that could be integrated in the proposed system. For example, our scheme could interoperate the *improved* SHIA approach, yielding a more efficient, $O(\log^2 n)$, successful aggregation.

Moreover, improvement can be achieved by changing the aggregation result from failure to success if ALS.II is executed. Or, including node values in the ALS.I confirmation message would allow the BS to obtain a partial view of the sensor field state. Moreover, ALS.I and ALS.II could be combined into a single phase, with benefits dependent on the nature of failures. Furthermore, the ATR protocol could be refined to improve its efficiency. To analyze such improvements (e.g., in the constants of the cost bounds), as well as factors such as cryptographic cost, simulations and experiments, will be part of our future work.

Finally, we discuss the exclusion of correct nodes. Thanks to the "onion" authentication of ALS.I, unlike the elementary approach, faulty nodes can at most modify confirmation messages of their children, but not those of other nodes in their subtree. However, the BS cannot distinguish between a child that provides an illegitimate message/inconsistent *ack* and a parent that modifies it. As a result, our scheme marks and then excludes both. In the worst case, $(\Delta-1)n_a$ correct nodes could be excluded. This bound could be improved to $n_a$, by allowing at most one child removal per parent and aggregation. This however would not necessarily guarantee a better performance on the average. Overall, the removal of correct nodes might eventually lead to a disconnected network: As they are in the proximity of faulty nodes, this could occur even if the adversary compromises *only partially* a node cut set. A less aggressive node exclusion could partially alleviate this: The BS maintains a "reputation" for all nodes, with marked nodes having their rating decreased and being excluded once they drop below a threshold. Moreover, ATR could be modified, so that marked and not yet excluded nodes are not neighbors in the new tree. The additional cost would be that the adversary disrupts more than $n_a$ sessions; evaluating this trade-off is also part of our future work.